\documentclass[dvipsnames,12pt,a4,epsfig]{article}
    \usepackage{xcolor}
  \usepackage{graphicx}
\newlength{\abstwidth}
\begin{document}
\newcommand{\flu}{fluctuations }
\newcommand{\hij }{{\footnotesize HIJING} }


\thispagestyle{empty}
\begin{center}
\noindent{\large {\bf Event by Event Analysis of High Multiplicity Events Produced in 158 A GeV/c $^{208}$Pb-$^{208}$Pb Collisions}}\\[5mm]
\end{center}

\noindent{Shakeel Ahmad$^1$\renewcommand{\thefootnote}{(\alph{footnote})}\footnote{email: Shakeel.Ahmad@cern.ch}, Anuj Chandra$^1$, Ashwini Kumar$^3$, {\color{Blue} O. S. K. Chaturvedi$^3$}, A. Ahmad$^2$, M. Zafar$^1$,  M. Irfan$^1$ and B. K. Singh$^3$\footnote{email: bksingh@bhu.ac.in}\\[5mm]
\noindent 1 {\footnotesize {\it Department of Physics, Aligarh Muslim University,Aligarh-202002,INDIA}}\\
\noindent 2 {\footnotesize {\it Department of Applied Physics, Aligarh Muslim University,Aligarh-202002,INDIA}}\\
\noindent 3 {\footnotesize {\it Department of Physics, Banaras Hindu University, Varanasi-221005,INDIA}}\\

\noindent {\footnotesize PACS Number(s): 29.40.Rg {Nuclear emulsions}}\\
\noindent {\footnotesize PACS Number(s): 25.75-q {Relativistic heavy-ion collisions}}\\
\noindent {\footnotesize PACS Number(s): 25.75.Gz {Fluctuations and Correlations}}\\

\noindent {\footnotesize {\bf Abstract}: An extensive analysis of individual high multiplicity events produced in 158 A $GeV/c$  \(^{208}\)Pb-\(^{208}\)Pb collisions is carried by adopting different methods to examine the anomalous behavior of these rare events. A method of selecting the events with densely populated narrow regions or spikes out of a given sample of collision events is discussed. Employing this approach two events with large spikes in their \(\eta\)- and \(\phi\)- distributions are selected for further analysis. For the sake of comparison, another two events which do not exhibit such spikes are simultaneously analyzed. The findings suggest that the systematic studies of particle density fluctuations in one- and two-dimensional phase-spaces and comparison with those obtained from the studies of correlation free Monte Carlo events, would be useful for identifying the events with large dynamical fluctuations. Formation of clusters or jet like phenomena in multihadronic final states in individual events is also discussed and the experimental findings are compared with the independent particle emission hypothesis by carrying out Monte Carlo simulations.}\\

\newpage
\section{Introduction}

The occurence of a phase transition between hadronic state and the quark-gluon plasma (QGP), have been a subject of immense importance in heavy-ion collision experiments. A key problem in this serach is the identification of signatures of QGP formation via the study of experimental obsrevables. The study of fluctuations and correlations have been suggested as a useful means for revealing the mechanism of particle production  along with some exotic phenomenon such as the {\color{Blue} possibility} of formation of Quark-gluon plasma (QGP) in heavy-ion collisions at high energies\cite{[cps],[ash],[bass]}. In heavy-ion collisions, as the system undergoes phase transition from hadronic matter to QGP, the degrees of freedom in two phases is quite different. Due to this large difference, correlations and fluctuations of the {\color{Blue} thermodynamic} quantities and/or the produced particle distributions in phase-space may change, apparently lacking any definite pattern. It is therefore necessary to analyse these collision processes on an event-by-event (ebe) basis. Therefore, an  analysis of high multiplicity events on ebe basis can be quite useful and can hint towards the exotic phenomenon along with the underlying physics process involved during the nucleus-nucleus (AA) collisions. In this search, any physical quantity measured in experiments is subjected to fluctuations. These fluctuations are believed to depend on the property of the system under study and thus expected to provide useful informations about the system formed during the collision\cite{1,2,62,3}. A non-monotonic evaluation of fluctuations observed in these collisions is suggested to serve as a signature for the phase transition at critical point. This, in turn, has generated considerable interest in the investigations involving fluctuations in AA collisions at SPS, RHIC and LHC energies. Several hadronic observables, produced in central \(^{208}\)Pb-\(^{208}\)Pb collisions  have been observed\cite{4} to exhibit qualitative changes in their energy dependence in the SPS energy range. A comparison of these observations with the predictions of statistical and (or) hadronic transport models,  indicates that the experimental results are consistent with the expected signals of the onset of a phase transition in AA collisions at SPS energies\cite{4,5}. Findings from {\color{Blue} the} Au-Au data at RHIC energies\cite{6,7} indicate that a highly collective and nearly thermalized system is formed in these collisions. Recent measurements on {\color{Blue} the} net-charge fluctuations at LHC energies too indicate their origin in the QGP phase and need dynamical model calculations to better understand the results\cite{[alice]}.  Furthermore, fluctuation observables are intrinsically related to particle correlations and will also be quite useful in providing some insights on the mechanism of particle production in heavy ion colisions\cite{[chai],[koch]} which is still an unresolved issue due to the inapplicability of perturbative QCD in soft regime\cite{[cps],[ash]}. Furthermore, the analysis of single event with large statistics can reveal very different physics than studying averages over a large statistical sample of events which becomes diluted or completely lost. In order to extract new physics associated with the fluctuations, it is quite necessary to understand the role of expected statistical fluctuations arising due to finiteness of the number of events. Information about the existence and nature of  phase transition may be obtained by investigating ebe fluctuations of suitable observables, like, transverse energy (\(E_t\)), transverse momentum (\(p_t\)), \(\ k/\pi \) ratio, electric charge, etc. Several attempts have been made\cite{7,8,9,10} to investigate ebe fluctuations using SPS and RHIC data in terms of these variables and the conclusions drawn are quite interesting. The main aim of all these investigations is to search for the rare events (such as due to QGP) exhibiting unusual behavior from a data sample with large statistics\cite{11,12,25,13,14,15}. 
Analysis of individual events produced in the collisions of heavy ions, like,  \(^{179}\)Au or \(^{208}\)Pb at SPS and higher energies have been argued\cite{11,12,25,13,14,15} to be statistically reliable as the multiplicities of such events {\color{Blue} are} high enough and the statistical fluctuations may be treated as under control. Furthermore, by studying the events with strong fluctuations, one can extract relevant informations on  the dynamical components[N17] present in comparison to the statistical component and can provide more insights into the underlying dynamics of high multiplicity events \cite{[flu19]}. Analysis of one 'hadron rich' event\cite{16} in the context of Centauro event, single event p$_t$ distribution\cite{171}, single event \(k/\pi\) ratio\cite{18}, intermittency in individual events\cite{1,18}, etc., are such studies undertaken so far.\\

An attempt is, therefore made to  analyze a few high multiplicity individual events produced in 158 A \(GeV/c\)  \(^{208}\)Pb-\(^{208}\)Pb interactions. These events are selected from EMU01 data collected using the conventional emulsion technique. The exposure was controlled by measuring heavily ionizing particles with a scintillator and discriminator with high threshold setting. The emulsions were exposed perpendicular to the $^{208}$Pb beam with momentum 158 A \(GeV/c\) in chambers of 20 emulsion plates. Each emulsion plate consisted of a thick acrylic base coated with a Fuji ET-7B emulsion layer on each side. The counter was placed behind the chamber which were in the beam during one 5 second pill. The spot size was about 6 cm\(^{2}\) corresponding a beam density of the beam about 5\(\times\)10\(^{2}\) nuclei/cm\(^2\). Based on {\color{Blue} the} parameterization of nucleus-nucleus inelastic cross-section {\color{Blue} ,} \(\sigma_{PbPb}\) is found to be 7.01 b\cite{sigma}. To select a sample of relatively central interactions, the emulsion plates directly below each target were visually scanned for high-multiplicity events. After the initial scanning selections were made, each event was examined in all the plates upstream of the interaction and rejected if the primary was noticeably less ionizing or if the primary had gone an additional interaction. The plates adjacent to the target allowed rejection of interactions occurring in emulsion rather than in the lead target. The event was also examined downstream and rejected if the remnants of the projectile {\color{Blue} were} present. Events with charge multiplicities above 1000 {\color{Blue} were} scanned efficiently, but those with lower multiplicities {\color{Blue} were} sampled incompletely. The emulsion experiment has an advantage over other detectors due to its 4\(\pi\) solid angle coverage and the data being free from biases as {\color{Blue} the} limited acceptance leads not only to the reduction of the charged particle multiplicity but may also distort to some of the event characteristics, {\color{Blue} such as the} particle density \flu\cite{12}. A comparative analysis is performed by generating \hij\cite{21} event samples with equal number of events as the experimental ones and with multiplicity distributions identical to the experimental ones. These events are simulated using the code \hij -1.35; other relevant details follow in the next sections. Furthermore, to test whether the fluctuations in some global observables characterizing an event are arising due to non-statistical reasons, distributions of various observables such as pseudo-rapidity distributions, azimuthal-angle distributions etc.  are compared with the corresponding reference distributions, obtained by {\color{Blue} the} event-mixing technique\cite{7,12}. In order to  minimize the contributions from the target and projectile spectators\cite{22} in the forward angle cone of \(\eta \ge\) 6.5, the charged particles having their \(\eta\) values lying in the range \(\eta_0 \pm 3\) have been considered for the analysis\cite{23}; \(\eta_0\) being beam rapidity (within 2.6-3.0). The values of  \(\eta_0\) are taken to be 3.5 and 3.2 respectively for the real and HIJING data sets.  A little shift of the peak will make no difference in the calculations/results.

\section{Results and Discussion}
\subsection{Presence of high density phase regions}
Even if the favorable conditions for the QGP formation are achieved, only few events are expected to be produced through the QGP. Therefore, the major task is to identify these rare events out of a large sample of {\color{Blue}the} collision events. An attempt is therefore made to find possible ways to characterize each event, which in turn, may lead to triggering of different classes of events and may identify the anomalous features. A search has also been carried{\color{Blue} out} for the high density phase regions in individual events where a lot of entropy is confined in a small domain. These high density regions are usually referred as the 'hot regions' or the {\color{Blue} spikes}\cite{12} and are searched in one-dimensional distributions on the basis of a quantity  \(d_{ik}\) \cite{12}, a  measure of the local deviation from the average particle density in units of statistical errors. For a particular distribution, values for the \(i^{th}\)-bin of the \(k^{th}\)-event are estimated using the prescription mentioned in ref.\cite{12}.\\
\noindent \(d_{ik}\) distributions in the {\color{Blue} pseudo-rapidity} ($\eta$) and azimuthal angle ($\phi$) spaces for 54  \(^{208}\)Pb-\(^{208}\)Pb events, each having \(N_k \ge\)  1000, are compared with the reference distributions i.e., 10 sets of mixed event samples as shown in Fig. 1. It should be mentioned that although the event-wise multiplicities are high enough $\sim$ (1000-1606), but the number of such events are only 54 and hence one may question whether the spiky regions observed in some of the events might be due to the statistical reasons. However, to ensure that the spikyness in these events are due to the reasons other than the statistical ones, ten sets of mixed event data, each set generated using a different sequence of random numbers as done in ref.{\color{Blue}\cite{8}} is considered. Analysis of these sets would, thus permit to put upper and lower limits of {\color{Blue} the} statistical fluctuations. Any significant deviation of the findings based on the real data from the ranges set by these mixed event samples would thus lead to the presence of dynamical contents. A marked difference in the \(d_{ik}\)-distributions for the experimental and the corresponding mixed-events is clearly observed in the form of large tails present in the regions of high \(d_{ik}\) values  which indicates that the real data do have a few events having spikes or hot regions in both $\eta$ and $\phi$ spaces defined as the regions with \(d_{ik} \ge 2.5\) in accordance with the ref.\cite{4}. In Table~1, the probability of occurrence of spikes, \(P(d_{ik})\) and the average size of spikes, \(<d_{ik}>\) with \(d_{ik} \ge 2.5\) for various data sets are presented for further analysis.  In order to compare the findings based on the real data with the predictions of Monte Carlo model \hij a parallel analysis of data samples generated using the code, \hij -1.35 and the corresponding mixed event samples has also been carried out. For this purpose; {\color{Blue} ten} sets of \hij events, each consisting of 10000 events, are simulated with impact parameter  between 0-6 $fm$.  From each of these sets, a set of 54 events having the  same multiplicity distributions as the experimental one are sorted out.  Further, ten sets of mixed (10000) events are then simulated to get ten data samples, each consisting of 54 mixed events and having multiplicity distributions identical to that for the experimental ones.
\(d_{ik}\)-distributions for these \hij and the corresponding mixed events are further analyzed which suggest that the distributions for all the sets acquire almost identical shapes, particularly in the region of large \(d_{ik}\) with \(d_{ik}\geq 2\) and appear to be similar to those due to the mixed event sets. The values of \(P(d_{ik})\) and \(<d_{ik}>\) for various bin widths in both \(\eta\) and \(\phi\) spaces are presented in Table 2. Based on the \(d_{ik}\) values, one can infer that although the occurrence of spikes is rare, but their presence can not be overlooked in the experimental data. These spikes are present in the real data but their presence is observed neither in the distributions generated using random number nor in the distributions generated by HIJING simulation. For the real data, the average size of spike, \(<d_{ik}>\) with \(d_{ik} \ge 2.5\) is observed to be larger than those for the mixed events. This difference in the \(d_{ik}\) values for the real and mixed events becomes more noticeable as the bin size, \(\Delta \eta\) or \(\Delta \phi\) becomes smaller. For the \hij events, \(d_{ik}\) values for the data and mixed events are found to be nearly the same.
\subsection{$\eta$ and $\phi$ -distributions of single event}
By studying the \(d_{ik}\)-distributions it may, therefore, be possible to identify the rare events having spikes. Using this criteria, two events which exhibit distinct spikes in their \(\eta\)- and \(\phi\)-distributions have been selected for further analysis and henceforth referred to as Event-1 and Event-2 with multiplicities 1606 and 1421, respectively. To check whether these spiky events exhibit some unusual characteristics, another two events with their \(\eta\)- and \(\phi\)-distributions nearly similar to those obtained for the entire sample are also analyzed and are referred to as Event-3 and Event-4 with multiplicities 1213 and 1469, respectively. In order to ensure that the observed fluctuations are the event characteristics and are not due to the statistical reasons, a parallel analysis of the events reproduced by {\color{Blue} the event-mixing technique} is also undertaken. For this purpose, from each of ten samples of mixed events, four events corresponding to the real ones are picked up for the analysis. Moreover, to test the presence of ebe fluctuations in \hij Monte Carlo Models,  four events from each of the ten samples of \hij and \hij -mixed events are selected and analyzed with multiplicities being the same as those of the experimental ones mentioned above so that comparative study may become more realistic and reliable to draw any firm conclusion.\\
In Fig. 2, \(\eta\)- and \(\phi\)-distributions of the four marked experimental events are compared with the mixed event sets. It is interesting to note that both \(\eta\)- and \(\phi\)-distributions for Event-1 and Event-2 show rather more pronounced peaks and valleys as compared to those arising in the case of mixed events.  On the other hand, For Events-3~and~-4, no such regions are observed and the magnitude of fluctuations in the particle densities appear to be rather smaller than those exhibited by the mixed event sets. These observations, thus indicate the presence of some 'hot regions' in the first two real events. Particle densities in these hot regions are found to be higher than the expected average value by a factor of \(\sim\) 2. Such inhomogeneity in pseudo-rapidity may arise either due to a very strong jet, i.e.,  large number of particles having their azimuthal and polar angle values very close to each other or due to several jets, each with rather smaller number of particles having similar values of polar angles but differ in azimuthal angles\cite{14}. Thus these observations further support that the significant fluctuations observed in Event-1 and Event-2 are mainly of the dynamical origin.  \\
To further corroborate our observations, \(\eta\)- and \(\phi\)-distributions for the five sets of \hij and the corresponding mixed events were analyzed. It was noticeable that both the distributions acquire almost similar shapes and no noticeable peaks/valleys were present in either \(\eta\)- or \(\phi\)-distribution of any individual events. Almost identical trends  have been observed in the remaining five sets of \hij events (not shown). Thus, the significant deviation of the distribution of the real data from the simulated data clearly hints towards the presence of the fluctuations with dynamical origin. 
\subsection{ Clusterization in individual events}Clusters are believed to be formed during the intermediate stage of an collision process, which finally decay into real physical particles. Their properties have been explained by the concept of “cluster” emission.  In this scenario, a great success has been achieved in describing many features of particle production in such collisions~\cite{[alver]}. The present analysis search for the presence of high density regions in two of the real events considered. Such regions of high particle density are envisaged to arise, due to the decay of either a heavy cluster or several clusters/jets of relatively smaller sizes\cite{12,14}. Therefore, to ensure further,  that the observed  spikiness in two of the events  might have some dynamical origin, an attempt is made to examine the  presence of cluster or jet-like phenomena following the algorithm applied to \(p\overline{p}\) data\cite{40}. This algorithm is somewhat different to that adopted in {\color{Blue} ref.~\cite{41}} in which formation of clusters and their sizes were looked into by histogramming the rapidity differences between the n$^{th}$ nearest neighbors. The present approach helps search for high density regions in \(\eta-\phi\) phase space which provides a clean separation in the \(\eta-\phi\) metric in the low multiplicity and low particle density final states\cite{12}. In order to test how the jet algorithm works for high density data and to what degree of clustering in the two-dimensional phase space one may expect, the analysis of the four real events is carried out. In these real events, the two-particle correlation can be studied in terms of the correlation length and extent in \(\eta-\phi\) phase space by the measured cluster multiplicity and/or the average number of particles in a cluster. Therefore, the method adopted here is expected to help estimating the cluster multiplicities and cluster frequencies on ebe basis. Since these observables are very sensitive to total event multiplicities, a comparison of the findings based on the real and simulated events with matching multiplicities is expected to lead to some definite conclusions. A detailed description about the method of analysis may be found in {\color{Blue} refs.~\cite{41}}.  Using this approach,  number of clusters (\(<m>\)) in each event with each cluster having at least \(m\) particles and number of particles  in each cluster are determined for a given value~of~\(r\) are determined. It may be mentioned here that for a very small value of \(r\) only a few or no clusters would be formed, whereas for a very large value of \(r\), almost all the particles would form one large cluster.\\

Results based on this algorithm for clusters with multiplicity, \(m \ge 5\) are discussed here for the four experimental and the corresponding mixed events. The variation of number of clusters, \(N_{cl}\) with \(r\) and the variation of mean cluster multiplicity, \(<m>\) with \(r\) {\color{Blue} are} shown in Fig. 3. A similar analysis is carried out for the \hij events and the corresponding mixed events too (not shown here). {Analysis presented here} provides an opportunity for a comparative study of the experimental data with such simulated events where the dominant sources of cluster formation are known. Based on the analysis, following observations are made:\\
{\underline{\it Mixed events}:}
Average cluster multiplicity (\(<m>\)), is observed to increase monotonically with increasing \(r\), from a minimal value of 5 to $\sim$ 15. The trends observed for all the ten sets of each of the four events are nearly the same irrespective of the event multiplicities which varies from  $\sim$ 1200 to 1600. Variations of \(N_{cl}\) with \(r\)  for all the events, too indicate almost one ('quiet') pattern having a broader maximum for the values of \(r\) between 0.2-0.3; the values of maximum number of clusters are found to be $\sim$ (120-130) and thereafter decreases gradually with increasing \(r\). Such broader maxima might be due to the randomness of these event structures\cite{42}.\\
{\underline{\it Real events}:}
It may be noted that the variation of \(N_{cl}\) with \(r\) for the two non-spiky events on {\color{Blue} qualitative} level are nearly the same to those due to the mixed events except that the maximum values of \(N_{cl}\) are slightly larger than those due to the mixed events. Experimental curve of \(<m>\) with \(r\) for 'Event-3' is found to overlap with the curve for simulated mixed events. For 'Event-4', \(<m>\) acquires somewhat larger values as compared to those due to the mixed events at \(r\) $\sim$ 0.6 and beyond. This deviation of the experimental curve with those for mixed events might be due to the presence of some clustering effects present in the real data.\\
Regarding the two spiky events, it is quite obvious that \(<m>\) grows with \(r\) faster than the one expected from the mixed event trends. It is interesting to see that this increase is much faster for 'Event-1' and can not be overlooked. The maximum value of the mean cluster multiplicity, in this case, is found to be $\sim$ 30, much higher than those due to the two non-spiky events or the mixed events. The variation of \(N_{cl}\) with \(r\) for the two spiky events are found with their maxima shifted towards the lower values of \(r\) and the subsequent decrease in the \(N_{cl}\) values with increasing \(r\) is much faster. This might be attributed to the idea\cite{42} that the majority of the produced particles in these events goes into a single or a few clusters.\\
{\underline{\it \hij events}:}
Dependence of \(<m>\)  and \(N_{cl}\) on \(r\) is found to exhibit almost identical trends for all ten sets of four events and is incidentally similar to the corresponding mixed events. The maximum value of \(<m>\) is $\sim$ 15, while the maximum number of clusters are $\sim$ (120-130) corresponding to \(r\) $\sim$ 0.3. Beyond this value, a slow decrease in the trend for the \(N_{cl}\) with \(r\) is observed. It may, therefore be pointed out that in the \hij events no predominant clustering effects are present.
Thus, these observations suggest that the clustering procedure adopted may help selecting events of special interest. Such procedures, if applied to the data with identified particles and known momenta, may lead to draw  even more interesting conclusions.
\section{Conclusions}  
High multiplicity events obtained from 158 A \(GeV/c\)  \(^{208}\)Pb-\(^{208}\)Pb collisions {\color{Blue} are} analysed {\color{Blue} in terms of} fluctuations in particle densities in pseudo-rapidity \& azimuthal angle space,  and the possibility of presence of mini-jets/jets {\color{Blue} are} looked for gaining insight to the origin of fluctuations observed in the experimental data. A comparative study is performed using the corresponding simulated events obtained from the generated mixed event and HIJING event samples. All these information have been put together in order to make any firm conclusion. Enhanced particle densities observed in narrow \(\eta\) and \(\phi\) bins  in the \(d_{ik}\) distributions are compared with the reference distributions due to the mixed events. The findings reveal the presence of few events in the real data sample with high density or spiky regions in \(\eta\) and \(\phi\) phase space. Such regions in the \hij events are also observed but with rather smaller magnitude.
Two spiky events out of real data sample suggest the presence of spikes in their \(\eta\) and \(\phi\) distributions where the particle densities are larger by a factor of $\sim$ 2. The two \hij events also exhibit similar densely populated regions but on rather smaller scale. On qualitative level, there is a noticeable difference between experimental and simulated events which suggest the presence of dynamical fluctuations in the spiky-events. A comparative analysis of the experimental results with those obtained from the simulated mixed events further suggests that the observed fluctuations in the particle densities of the two spiky events might have dynamical contents, and are promisingly suitable for the analysis. A dominant cluster like phenomena in two dimensional \(\eta\)-\(\phi\) phase space is observed for  two real events (spiky events) having densely populated regions. This very dominance of clusterization is observed to disappear in the case of mixed events. Findings of the comparative analysis of real events with \hij events indicate that the clustering effects are significantly stronger in the case of real events selected for the study. The experimental findings do not agree with the hypothesis of completely independent particle emission as the results {\color{Blue} indicate} towards the cluster like phenomenon present in the real events. The Bose-Einstein correlation, contributes to the origin of cluster formation due to the correlated emission of identical mesons so can not be ruled out as one of the possible sources of large local fluctuations observed in the particle densities. This effect is of quantum statistical origin and is not incorporated in the simulation framework of a transport model HIJING used here. The differences observed in the experimental data with those of the simulation results can also be interpreted in terms of certain nontrivial dynamics involved such as collective flow. Therefore, {\color{Blue} further} investigations are much needed and motivates us for future analysis of the available experimental data.
In conclusion,  a comparative study of the real data with the simulated events highlights the underlying particle production mechanism and satisfactorily explains the observed features of experimental data available at 158 A  $GeV/c$. Analyses involving particle densities or clusterization in one and two dimensional phase space, although may not lead to some definite conclusions due to limited statistics yet it may be remarked that the methods adopted, may be used as a tool to select special events signaling the formation of some exotic states, eg. QGP and/or DCC and can contribute to the ongoing search. 
\section{Acknowledgments}
\noindent The authors thank to Prof. A. Bhasin, Jammu University, Jammu for providing 158 A $GeV/c$ data and also for the fruitful discussions during the compilation of this work.

\newpage
\noindent Table~1: Probability \%, P(\(d_{ik}\))  and  mean values, \(<d_{ik}>\) of the spikes with \(d_{ik} \ge 2.5\) for the  experimental and the corresponding mixed event samples (only minimum and maximum values from the ten sets of mixed events are presented).
\begin{footnotesize}
\begin{table}
\begin{center}
\begin{tabular}{|c|clll|}\hline 
   & & Expt. & \multicolumn{2}{c|}{Mixed Evt.} \\ 
   & &       & \multicolumn{2}{c|}{{\footnotesize(10 sets)}}  \\ \cline{4-5}
\(\Delta \eta\) & & & min. & max. \\
0.1 & P(\(d_{ik}\)) & 1.52 &  0.66 &  1.02  \\  
 & \(<d_{ik}>\) & 3.24$\pm$0.50  & 2.61$\pm$0.30 & 2.92$\pm$0.36\\ \hline 
0.2 & P(\(d_{ik}\)) & 2.57 & 1.05 & 1.58  \\
 & \(<d_{ik}>\) & 3.67$\pm$0.98  & 3.05$\pm$0.63 & 3.44$\pm$0.64\\ \hline 
0.3 & P(\(d_{ik}\)) & 3.44 & 1.26 & 1.96  \\
 & \(<d_{ik}>\) & 3.98$\pm$1.18  & 3.35$\pm$0.07 & 3.77$\pm$0.48\\  \hline
 &  \multicolumn{4}{c|}{}   \\[-2mm]
\(\Delta \phi\) &  \multicolumn{4}{c|}{}   \\ \hline
5$^O$ & P(\(d_{ik}\)) & 0.87 &  0.04 & 0.09 \\
 & \(<d_{ik}>\) & 3.09$\pm$0.53  & 2.56$\pm$0.14 & 2.97$\pm$0.31\\ \hline 
10$^O$ & P(\(d_{ik}\)) & 1.47 & 0.16 & 0.57 \\
 & \(<d_{ik}>\) & 3.55$\pm$0.71  & {\color{Blue}{2.70$\pm$0.01}} & 3.13$\pm$0.15\\ \hline 
15$^O$ & P(\(d_{ik}\)) & 1.92 & 0.92 & 1.34  \\
 & \(<d_{ik}>\) & 3.73$\pm$0.98  & 2.61$\pm$0.04 & 2.85$\pm$0.25\\  \hline 
\end{tabular}
\end{center}
\end{table}
\end{footnotesize} 

 \newpage
\noindent Table~2: Minimum and Maximum values of \(P(d_{ik})\)(\%) and \(<d_{ik}>\) with \(d_{ik} \ge 2.5\) out of ten sets of \hij and the corresponding mixed event sets.
\begin{footnotesize}
\begin{table}
\begin{center}
\begin{tabular}{|c|lllll|} \hline 
   & & \multicolumn{2}{c}{HIJING} & \multicolumn{2}{c|}{HIJING. Mixed}\\ 
   & & \multicolumn{2}{c}{\footnotesize(10 sets)} & \multicolumn{2}{c|}{{\footnotesize(10 sets)}}  \\ \cline{3-4}\cline{5-6}
\(\Delta \eta\) & & min & max & min & max \\ \hline
0.1 & P(\(d_{ik}\)) & 0.11 & 0.26 & 0.11 & 0.24 \\
 & \(<d_{ik}>\) & 2.61$\pm$0.13 & 2.83$\pm$0.26 & 2.67$\pm$0.07 & 2.93$\pm$0.22 \\ \hline 
0.2 & P(\(d_{ik}\)) & 0.11 & 0.63 & 0.12 & 0.32 \\
 & \(<d_{ik}>\) & 2.51$\pm$0.14 & 2.97$\pm$0.18 &2.63$\pm$0.14 & 2.98$\pm$0.18\\ \hline 
0.3 & P(\(d_{ik}\)) & 0.16 & 0.48 & 0.16 & 0.47 \\
 & \(<d_{ik}>\) & 2.59$\pm$0.17 & 3.03$\pm$0.15 & 2.62$\pm$0.18 & 2.74$\pm$0.16\\ \hline 
\(\Delta \phi\) &  \multicolumn{5}{c|}{}   \\ \hline
5$^O$ & P(\(d_{ik}\)) & 0.04 & 0.37 & 0.05 & 0.24 \\ 
 & \(<d_{ik}>\) & 2.63$\pm$0.20 & 2.93$\pm$0.19 & 2.57$\pm$0.14 & 2.96$\pm$0.19\\ \hline 
10$^O$ & P(\(d_{ik}\)) & 0.09 & 0.46 & 0.09 & 0.35 \\
 & \(<d_{ik}>\) & 2.56$\pm$0.26  & 3.00$\pm$0.41 & 2.56$\pm$0.22 & 3.30$\pm$0.23\\ \hline 
15$^O$ & P(\(d_{ik}\)) & 0.14 & 0.54 & 0.15 & 0.43\\
 & \(<d_{ik}>\) & 2.52$\pm$.16 & 2.93$\pm$.26 & 2.57$\pm$0.18 & 3.07$\pm$0.13\\ \hline 
\end{tabular}
\end{center}
\end{table}
\end{footnotesize} 
\newpage
\begin{figure}
  \begin{center}
 \includegraphics[height=10cm , width=12cm]{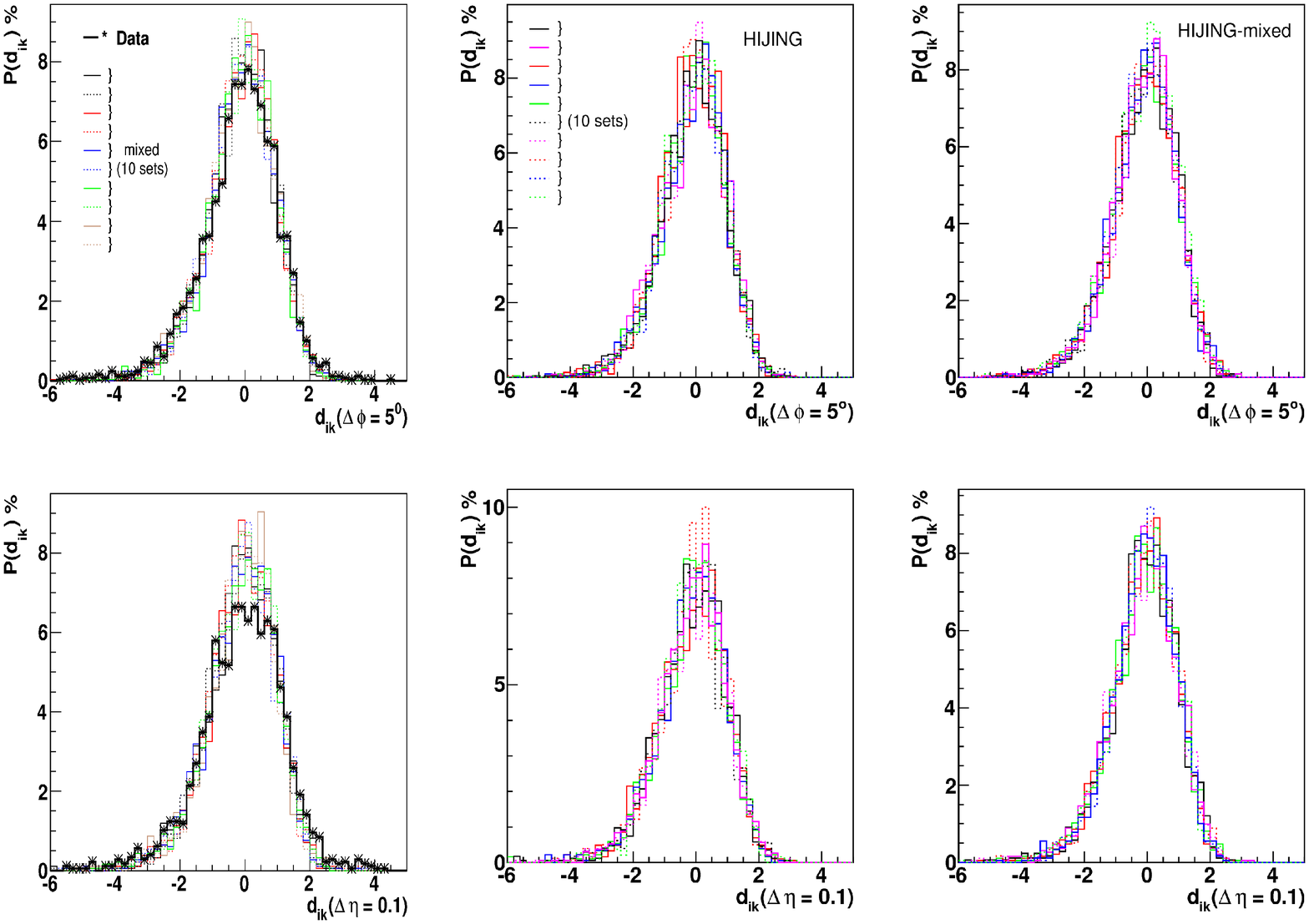}
\caption{$(d_{ik}$) distributions for the experimental events in pseudo-rapidity bins, \(\Delta \eta = 0.1\) and azimuthal angle bins, \(\Delta \phi = 5^0\) compared with  the  ten sets of mixed event samples (left), HIJING (middle) and HIJING-mixed (right) events.}
\label{fig1}
\end{center}
 \end{figure}
 
\newpage
\begin{figure}
  \begin{center}
 \includegraphics[height=8cm , width=12cm]{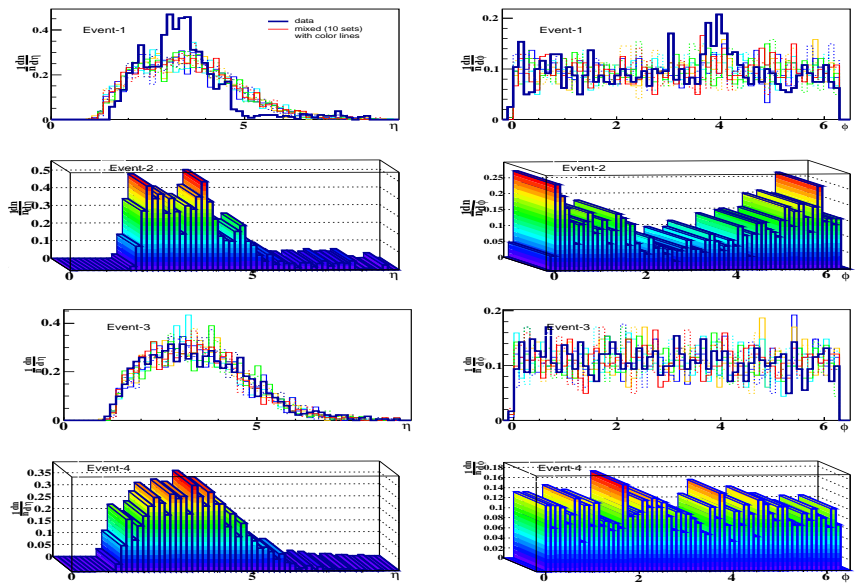}
\caption{\(\eta\) and \(\phi\) distributions of relativistic charged particles for the two spiky ({\it from the top}) and two non-spiky experimental events ({\it from the bottom}) with the mixed events.}
\label{fig2}
\end{center}
 \end{figure}
\newpage
\begin{figure}
  \begin{center}
 \includegraphics[height=8cm , width=12cm]{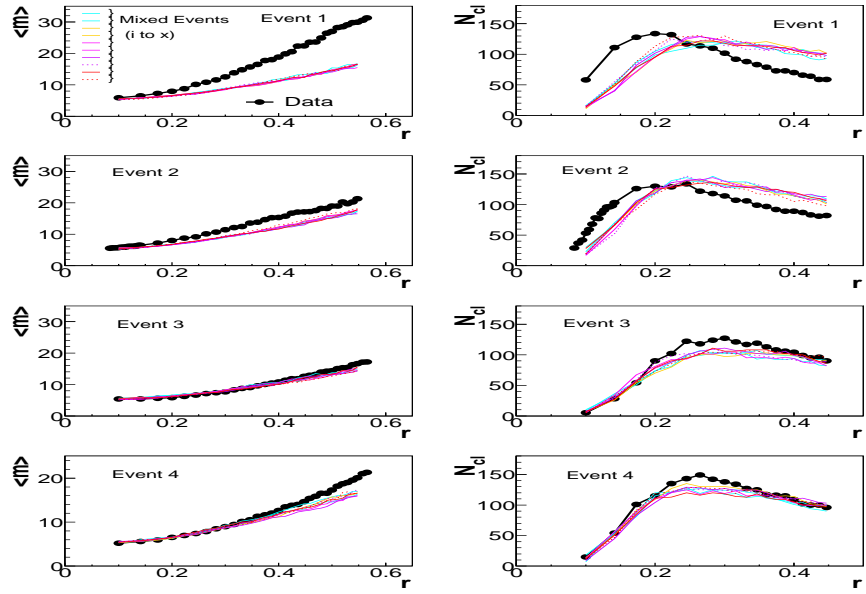}
\caption{Dependence of average cluster multiplicity, \(<m>\) and number of clusters, \(N_{cl}\) on the distance measure, \(r\) for the real  and mixed events.}
\label{fig3}
\end{center}
 \end{figure}

\end{document}